\documentclass[sigconf,screen,10pt,natbib=false,authorversion]{acmart}
\usepackage[abbreviations]{foreign}
\AtBeginDocument{\providecommand\BibTeX{{\normalfont B\kern-0.5em{\scshape i\kern-0.25em b}\kern-0.8em\TeX}}}

\usepackage[utf8]{inputenc}
\usepackage[T1]{fontenc}
\usepackage{xcolor}
\usepackage{amsmath,amsfonts}
\usepackage{graphicx}
\usepackage[inline]{enumitem}
\usepackage{subcaption}
\usepackage{tabularx, makecell, booktabs}
\usepackage{siunitx}
\usepackage[backend=biber,style=acmnumeric,minnames=1,maxnames=1,giveninits=true]{biblatex}

\usepackage{xpatch}

\makeatletter
\patchcmd\blx@bblinput{\blx@blxinit}
                      {\blx@blxinit
                       \def\jobname{paper}%new jobname
                      }{}{\fail}
\makeatother
			     
\AtBeginBibliography{\footnotesize}
\AtEveryBibitem{\clearfield{pages}\clearlist{location}\clearlist{organization}\clearfield{pagetotal}\clearfield{eprint}\clearfield{doi}\clearfield{isbn}\clearfield{number}\clearfield{volume}\clearfield{note}\clearfield{issn}\clearfield{series}\clearfield{editor}\clearfield{numpages}\clearlist{publisher}\clearlist{editor}\ifentrytype{online}{}{\clearfield{url}}\ifentrytype{inproceedings}{\clearfield{year}}{}}
\renewbibmacro*{year}{\iffieldundef{year}{}{\printfield{year}}}

\makeatletter
\@ifundefined{qty}{\newcommand*{\qty}[2]{\SI{#1}{#2}}\DeclareBinaryPrefix \kibi { Ki } { 10 }\DeclareBinaryPrefix \mebi { Mi } { 20 }\DeclareBinaryPrefix \gibi { Gi } { 30 }\DeclareBinaryPrefix \tebi { Ti } { 40 }\DeclareBinaryPrefix \pebi { Pi } { 50 }\DeclareBinaryPrefix \exbi { Ei } { 60 }\DeclareBinaryPrefix \zebi { Zi } { 70 }\DeclareBinaryPrefix \yobi { Yi } { 80 }\DeclareSIUnit \bit  { bit }\DeclareSIUnit \byte { B }}{}
\makeatother

\acmYear{2023}\copyrightyear{2023}
\acmConference[DEBS '23]{The 17th ACM International Conference on Distributed and Event-based Systems}{June 27--30, 2023}{Neuchatel, Switzerland}
\acmBooktitle{The 17th ACM International Conference on Distributed and Event-based Systems (DEBS '23), June 27--30, 2023, Neuchatel, Switzerland}
\acmPrice{15.00}
\acmDOI{10.1145/3583678.3596899}
\acmISBN{979-8-4007-0122-1/23/06}

\newbool{showcomments}
\setbool{showcomments}{true}
\ifthenelse{\boolean{showcomments}}
{ \newcommand{\mynote}[3]{
    \fbox{\bfseries\sffamily\scriptsize#1}
    {\small$\blacktriangleright$\textsf{\emph{\color{#3}{#2}}}$\blacktriangleleft$}}}
{ \newcommand{\mynote}[3]{}}

\usepackage[inline]{enumitem}
\setlist{noitemsep,topsep=0pt,parsep=0pt,partopsep=0pt}
\usepackage{setspace}
\usepackage[margin=5pt,font={stretch=0.9}]{caption}

\usepackage{titlesec}
\titlespacing*{\section}{0pt}{0.5\baselineskip}{0.2\baselineskip}
\titlespacing*{\subsection}{0pt}{0.5\baselineskip}{0.2\baselineskip}

\def\iphoneproFourteenEncSpeedup{13.9}
\def\iphoneproFourteenDecSpeedup{5.1}
\def\pixelproSevenEncSpeedup{6.9}

\def\iphoneproFourteenEncWasmMsMaxSize{307.2}
\def\iphoneproFourteenEncSjclMsMaxSize{4305.6}
\def\pixelproSevenFourteenEncWasmMsMaxSize{786.9}
\def\pixelproSevenFourteenEncSjclMsMaxSize{5700.4}
\def\encWasmMsMaxSizeRatio{2.6}
\def\encSjclMsMaxSizeRatio{1.3}
\def\maxSpeedup{14}
 

\makeatletter
\begin{document}

\title[Preventing EFail Attacks with Client-Side WebAssembly]{Preventing EFail Attacks with Client-Side WebAssembly: The Case of Swiss Post's IncaMail}
\subtitle{(Industry and Application Track)}

\author{Pascal Gerig}
\orcid{0000-0001-7826-9489}
\affiliation{\institution{University of Bern}
	\city{Bern}
	\country{Switzerland}
}
\author{Jämes Ménétrey}
\orcid{0000-0003-2470-2827}
\affiliation{\institution{University of Neuchâtel}
	\city{Neuchâtel}
	\country{Switzerland}
}
\author{Baptiste Lanoix}
\orcid{0009-0001-1736-0315}
\affiliation{\institution{Swiss Post}
	\city{Neuchâtel}
	\country{Switzerland}
}
\author{Florian Stoller}
\orcid{0009-0007-8417-2449}
\affiliation{\institution{Swiss Post}
	\city{Neuchâtel}
	\country{Switzerland}
}
\author{Pascal Felber}
\orcid{0000-0003-1574-6721}
\affiliation{\institution{University of Neuchâtel}
	\city{Neuchâtel}
	\country{Switzerland}
}
\author{Marcelo Pasin}
\orcid{0000-0002-3064-5315}
\affiliation{\institution{Haute École Arc, HES-SO}
	\city{Neuchâtel}
	\country{Switzerland}
}
\author{Valerio Schiavoni}
\orcid{0000-0003-1493-6603}
\affiliation{\institution{University of Neuchâtel}
	\city{Neuchâtel}
	\country{Switzerland}
}

\renewcommand{\shortauthors}{Gerig P. et al.}

\begin{abstract}	
Traditional email encryption schemes are vulnerable to EFail attacks, which exploit the lack of message authentication by manipulating ciphertexts and exfiltrating plaintext via HTML backchannels.
Swiss Post's IncaMail, a secure email service for transmitting legally binding, encrypted, and verifiable emails, counters EFail attacks using an \emph{authenticated-encryption with associated data} (AEAD) encryption scheme to ensure message privacy and authentication between servers.
IncaMail relies on a trusted infrastructure backend and encrypts messages per user policy.
This paper presents a revised IncaMail architecture that offloads the majority of cryptographic operations to clients, offering benefits such as reduced computational load and energy footprint, relaxed trust assumptions, and per-message encryption key policies.
Our proof-of-concept prototype and benchmarks demonstrate the robustness of the proposed scheme, with client-side WebAssembly-based cryptographic operations yielding significant performance improvements (up to \textasciitilde\maxSpeedup$\times$) over conventional JavaScript implementations.
\end{abstract}



\keywords{EFail, Email, WebAssembly, Cryptography, Mitigation}

\maketitle

\section{Introduction}\label{s:intro}

Secure messaging has become increasingly important in today's digital landscape, where privacy and security are paramount concerns.
Encryption schemes are critical in ensuring the confidentiality and integrity of messages transmitted over untrusted networks.
However, traditional email encryption approaches suffer from various vulnerabilities.
EFail attacks~\cite{efail2018} exploit the lack of message authentication in encrypted emails to manipulate ciphertexts and exfiltrate plaintext data through maliciously crafted HTML backchannels.
To counter such threats and enhance the security of email communication, Swiss Post introduced IncaMail~\cite{incamail}, a secure email service designed for transmitting legally binding, encrypted, and verifiable emails.
IncaMail employs an \emph{authenticated encryption with associated data} (AEAD) encryption scheme to guarantee message privacy and authentication between servers.
The current architecture offers a robust solution against EFail attacks.
However, it also presents certain limitations due to its centralised nature, such as high computational load on the server, sustained network traffic between clients and servers, and the need for a fully trusted server that handles plaintext messages.

This paper presents an experimental IncaMail architecture revision that offloads most cryptographic operations to clients' browsers using WebAssembly (Wasm)~\cite{10.1145/3062341.3062363}, an innovative bytecode format for portable and efficient code execution in web applications.
Our proof-of-concept prototype fully embeds the industry-standard OpenSSL cryptographic library, compiled in Wasm and embedded in the web browser.
Our experimental evaluation shows significant performance improvements over alternative approaches based on JavaScript, while preserving linear time complexity with respect to message and file attachment size.
Client-side encryption offers numerous performance and architectural benefits, including reduced computational load and therefore lower energy footprint, relaxed trust assumptions, and per-message encryption key policies.
This work contributes to the advancement of secure email communication services and highlights the potential of Wasm to enhance the performance and security of such systems. \section{Background}\label{s:back}
In this section, we first provide background notions on WebAssembly (\S\ref{ssec:wasm}), subsequently delving into specifics of the EFail attacks, as well as existing countermeasures (\S\ref{ssec:efail}).

\subsection{WebAssembly Primer}\label{ssec:wasm}
WebAssembly (Wasm)~\cite{10.1145/3062341.3062363} is a portable bytecode format designed to provide optimal performance when compiled from various programming languages.
Wasm's efficient binary format and streamlined execution yield near-native performance and faster execution times for web applications.
It facilitates the integration of libraries and functionalities from other platforms into web applications, eliminating the need for reimplementation.
Its platform-agnostic nature enables code to be written once and executed across various operating systems and browsers.
Wasm supports compilation from multiple programming languages, such as C, C++, Rust, and more, granting developers to leverage their existing expertise and codebases, employing compiler infrastructures such as LLVM~\cite{10.5555/977395.977673}.
Additionally, Wasm supports multithreading and SIMD instructions, enabling developers to take advantage of modern hardware capabilities, optimising performance for tasks like cryptographic operations.
Lastly, Wasm's sandboxed execution environment mitigates security vulnerabilities by isolating them from the underlying system.

\textbf{Wasm in web browsers.}
Wasm is supported across all modern web browsers and serves as a complementary solution to JavaScript rather than a replacement.
Wasm lacks access to contextual web page information, \eg the Document Object Model (DOM), and instead focuses on providing efficient computations for CPU-intensive tasks.
This complementary relationship allows developers to seamlessly integrate Wasm with JavaScript, allowing for the incremental adoption of Wasm and leveraging the strengths of both technologies.
Emscripten~\cite{10.1145/2048147.2048224}, based on the LLVM toolchain, offers a higher-level compiler toolchain for web application development.
Beyond compiling Wasm code, Emscripten facilitates seamless interaction between Wasm functions and JavaScript, including function calls and data transfers.

\subsection{EFail: attacks and counter-measures}\label{ssec:efail}
EFail~\cite{efail2018}, a security vulnerability disclosed in 2018, compromises the confidentiality of email encryption protocols, specifically OpenPGP and S/MIME, by exploiting implementation flaws and allowing adversaries to decrypt messages.

\textbf{EFail Attack Overview.}
The EFail attack exploits weaknesses in the interaction between email clients and encryption plugins, as well as the properties of the encrypted messages themselves.
The attack primarily targets HTML-rendering email clients that utilise OpenPGP or S/MIME encryption.
Two distinct variants of the attack exist: the \emph{Direct Exfiltration} attack and the \emph{CBC/CFB Gadget} attack.
Both variants leverage the concept of malleability in encrypted messages, allowing attackers to manipulate ciphertext without knowing the corresponding plaintext.
In the Direct Exfiltration attack, the adversary modifies the encrypted email by injecting an image tag with a crafted URL that will contain the decrypted content once the email client processes the modified email.
When the victim's email client decrypts the message and loads the injected image, the plaintext message is inadvertently sent to the attacker-controlled server.
The CBC/CFB Gadget attack is more sophisticated and relies on manipulating the block cipher modes of operation in S/MIME and OpenPGP.
By carefully crafting the ciphertext and exploiting the lack of integrity protection, the attacker induces specific plaintext patterns, which, combined with HTML tags, can be used to exfiltrate the decrypted content in a similar manner to the Direct Exfiltration attack.

\textbf{Countermeasures.} The EFail attack threat can be mitigated with countermeasures at the email client and protocol levels.
These include: 
\emph{(1)} disabling HTML rendering: since the EFail attack relies on the HTML rendering capabilities of email clients to exfiltrate decrypted data, disabling HTML rendering in the client settings can effectively prevent the attack; 
\emph{(2)} integrity protection: implementing cryptographic mechanisms, such as authenticated encryption (AEAD) or message authentication codes (MAC), can ensure the integrity of encrypted messages and thwart attempts to manipulate ciphertexts; and 
\emph{(3)} secure implementations: email client developers and encryption plugin providers should adhere to best practices for implementing encryption protocols, such as proper handling of decryption errors and avoiding leakage of plaintext data through external resources.

The EFail attack highlights the importance of proper implementation and usage of encryption protocols in email communications.
This paper shows how authenticated encryption at the client-side browser effectively addresses the risks associated with EFail, while safeguarding the ongoing confidentiality of encrypted email communications.
 \section{Related Work}\label{s:relw}
In this section, we cover existing approaches to mitigate \textsc{EFail} attacks (\S\ref{ssec:rw:efail}), how Wasm is used to mitigate other security attacks (\S\ref{ssec:rw:wasmmitig}), and the current state of Wasm deployments in the wild for real-world application scenarios~(\S\ref{ssec:rw:wasmwild}).
To our knowledge, we are the first to leverage WebAssembly in the context of efficient and secure web-based email clients.

\subsection{Mitigations to EFail attacks}\label{ssec:rw:efail}
In~\cite{schwenk2020mitigation}, authors propose generic mitigations against different variants of EFail attacks, such as \textsc{Reply}~\cite{katz2000chosen,muller2019re}, \textsc{EFail-Md} or \textsc{Efail-Mg}~\cite{efail2018}.
They do so by checking the decryption context, \eg the \texttt{SMTP} headers and \texttt{MIME} structure during decryption, implementing their solution into the Thunderbird email client and OpenPGP with support for AEAD. 
Our solution is not bound to a particular email client, and rather it is designed for web-based senders and receivers, potentially via mobile devices.
Further, we aim to provide an Outlook add-in as an extension of our approach.

\subsection{Wasm as general protection technique}\label{ssec:rw:wasmmitig}
WebAssembly allows developers to integrate shielding mechanisms into their large codebase, by leveraging Wasm's sandboxing approach for features such as memory isolation~\cite{lehmann2020everything}.
Wasm can also be formally verified~\cite{protzenko2019formally}.
However, given the native execution speed of Wasm binaries~\cite{wang2021empowering}, cybercriminals continuously try to abuse Wasm binaries to exfiltrate sensitive data.
Solutions exist to deploy a double-sandbox approach to fully isolate the Wasm runtime from the host and vice versa, for instance, leveraging off-the-shelf trusted execution environments~\cite{menetrey2021twine,menetrey2022watz}.
To integrate such TEE-based approaches in our revised IncaMail architecture, a TEE-aware abstraction layer available across a variety of devices (\ie, server, client, mobile) is currently missing. 

\subsection{Wasm in the wild}\label{ssec:rw:wasmwild}
WebAssembly has increasingly been adopted by industry, across several domains and contexts. 
Its native execution speed and cross-platform portability facilitate its adoption in the so-called cloud-edge continuum~\cite{menetrey2022webassembly}, spanning IoT devices up to server-grade deployments.
Wasm is the execution runtime for several classes of systems, \eg energy-efficient blockchain systems~\cite{dfinity2022internet}, serverless platforms~\cite{gadepalli2020sledge,gackstatter2022pushing}.

 \section{Use-Case: IncaMail}\label{s:inca}
The Swiss Post developed IncaMail, a service offering a secure way of transmitting information via emails. Crucially, this communication channel offers authenticated encryption which mitigates the EFail attacks.
Users can send messages using three input channels:
\emph{(1)} a web interface; 
\emph{(2)} an Outlook add-in; and 
\emph{(3)} a dedicated web API.

\textbf{Current workflow.} When a customer of IncaMail sends a secure email, the secure message is transmitted to IncaMail's backend servers for encryption using a symmetric cipher (\ie, AES).
The message is encrypted for each recipient.
The ciphertexts are embedded within standard emails as an attachment, and subsequently sent individually to the respective recipients.
The encryption keys are retained on IncaMail's premises, while the secure messages are not stored within the Swiss Post infrastructure, except in some instances for caching purposes.
Upon reception of the email, one or more recipients open the attachment.
This attachment is an HTML file containing an HTML form with predefined hidden fields, which include the ciphertext, the MAC, and other associated data.
A given recipient submits the form to the IncaMail server, which internally retrieves the encryption key, decrypts the submitted information, and displays the plaintext of the secure message.
Additionally, secure messages may also include secure file attachments, which are processed with a similar encryption mechanism.
This centralised architecture has the following limitations:
\begin{itemize}[noitemsep,topsep=0pt,leftmargin=*]
    \item all cryptographic operations happen solely on the server, demanding additional computing resources of Swiss Post than if the operations would take place on the client premises;
    \item since all messages and corresponding ciphertexts are transferred from and to Swiss Post infrastructure, this approach induces more network traffic than if the messages were decrypted on the client;
    \item the IncaMail server must be a fully trusted entity, as all messages are seen in plaintext. In case of errors or compromise, it could read and manipulate all messages;
    \item since all messages are encrypted with a private key bound per recipient, group messages are encrypted several times (one per each recipient).
\end{itemize}

\subsection{Offloading Cryptographic Operations}
In this work, we evaluate a revised architecture where some cryptographic operations are offloaded to the customers' browsers.
This offers many advantages over the current architecture, notably \emph{(1)} reducing the CPU and volatile memory usage, as the server no longer needs to encrypt and decrypt the secure message, and \emph{(2)} reducing the trusted computing base (TCB), as only the endpoint that receives and delivers encryption keys must be trusted, since plaintexts are no longer sent to the IncaMail servers.
Furthermore, we improved the encryption scheme of IncaMail to share a common encryption key for all the recipients of a given message, which would save non-volatile memory of IncaMail infrastructure.
\begin{figure}[!t]
    \centering
    \includegraphics[width=\columnwidth]{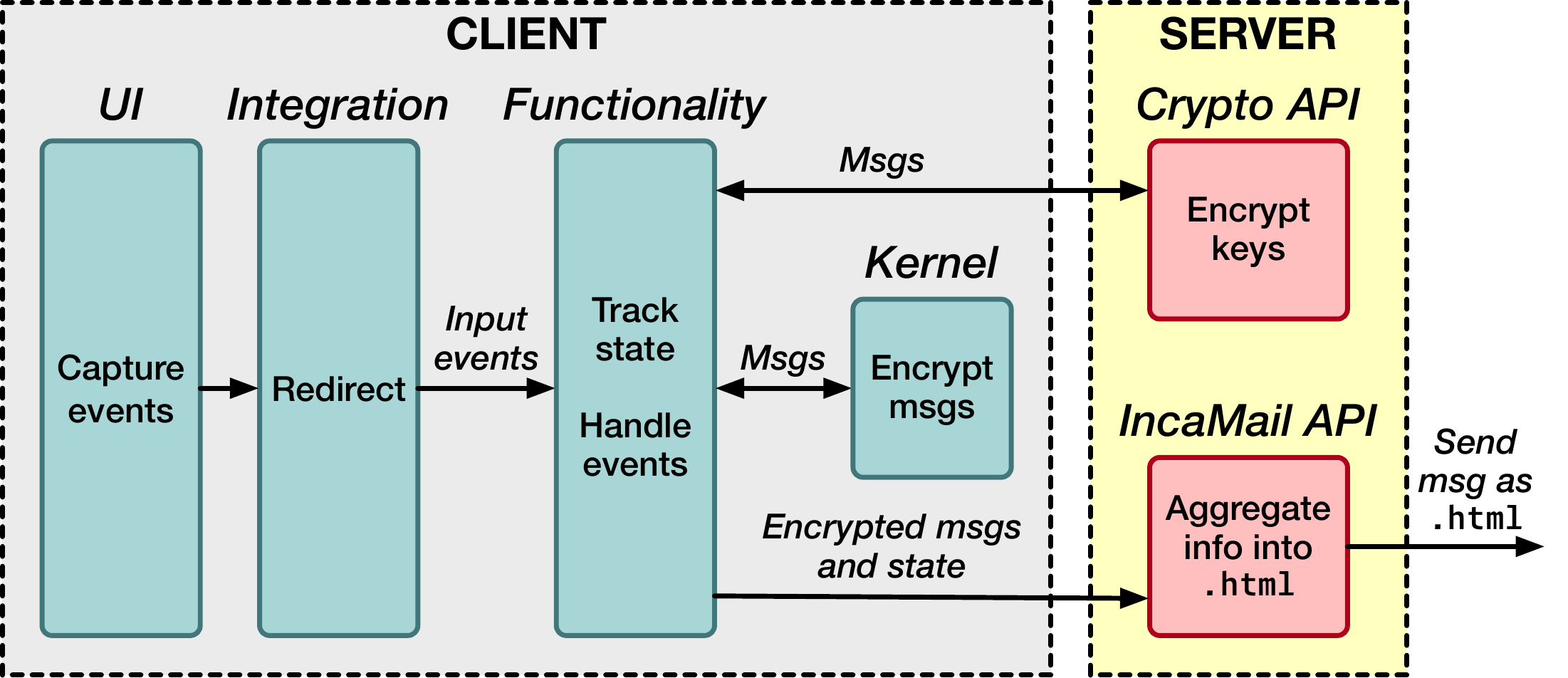}
    \caption{Workflow of sending secure messages.}
    \label{fig:sending}
\end{figure}

\textbf{Wasm as a generic solution.}
Although the W3C's Web Cryptography API~\cite{10.1145/2567948.2579224} facilitates seamless access to various cryptographic operations in modern browsers, we opted for Wasm as a comprehensive solution for managing cryptographic computations, employing OpenSSL~\cite{openssl} as the cryptography library, due to a number of considerations.
Firstly, Swiss Post maintains numerous technologies for servicing IncaMail (\eg web interface, Outlook add-in).
Wasm facilitates the use of well-established cryptography libraries across various platforms and browsers, ensuring compatibility even when the native Web Cryptography API may not support specific algorithms.
As OpenSSL maintains a consistent API across platforms, developers can leverage existing code and knowledge when implementing cryptographic operations.
This consistency streamlines development and maintenance in comparison to using different APIs for different platforms.
Secondly, Wasm is designed for efficiency and speed.
While it may not achieve the performance of native code, it significantly surpasses JavaScript in terms of execution speed.
This paper demonstrates that employing OpenSSL yields superior performance compared to JavaScript-based alternatives.
Lastly, OpenSSL offers greater control and flexibility than the Web Cryptography API, as it grants developers access to a broader range of cryptographic primitives and fine-tunes their implementations to satisfy specific requirements.
This advantage frees Swiss Post from being constrained by the choices of browser implementations.

\textbf{OpenSSL compilation.}
A challenge we faced was the compilation of OpenSSL, which is required by the absence of Wasm-compiled binaries in the official OpenSSL distribution.
For that purpose, we used Emscripten and disabled certain features, including hardware acceleration due to the inaccessibility of assembly instructions in Wasm, and multithreading, given their irrelevance to our use case.

\textbf{Architecture.}
We adapted the IncaMail architecture by implementing several modifications.
The key generation operations are now delegated to the client side, while a new server-side RESTful API is responsible for managing the keys.
This API, \ie \emph{Crypto API}, serves as the trusted computing base (TCB) for IncaMail.
The encryption and decryption of secure messages and file attachments are also offloaded to the client side, ensuring that the plaintext of messages no longer passes through the IncaMail infrastructure.
Consequently, we have improved the threat model of the IncaMail backend infrastructure, allowing it to adhere to an honest-but-curious model in which the system can inspect exchanged information without compromising the confidentiality of messages.

\begin{figure}[!t]
    \centering
    \includegraphics[width=\columnwidth]{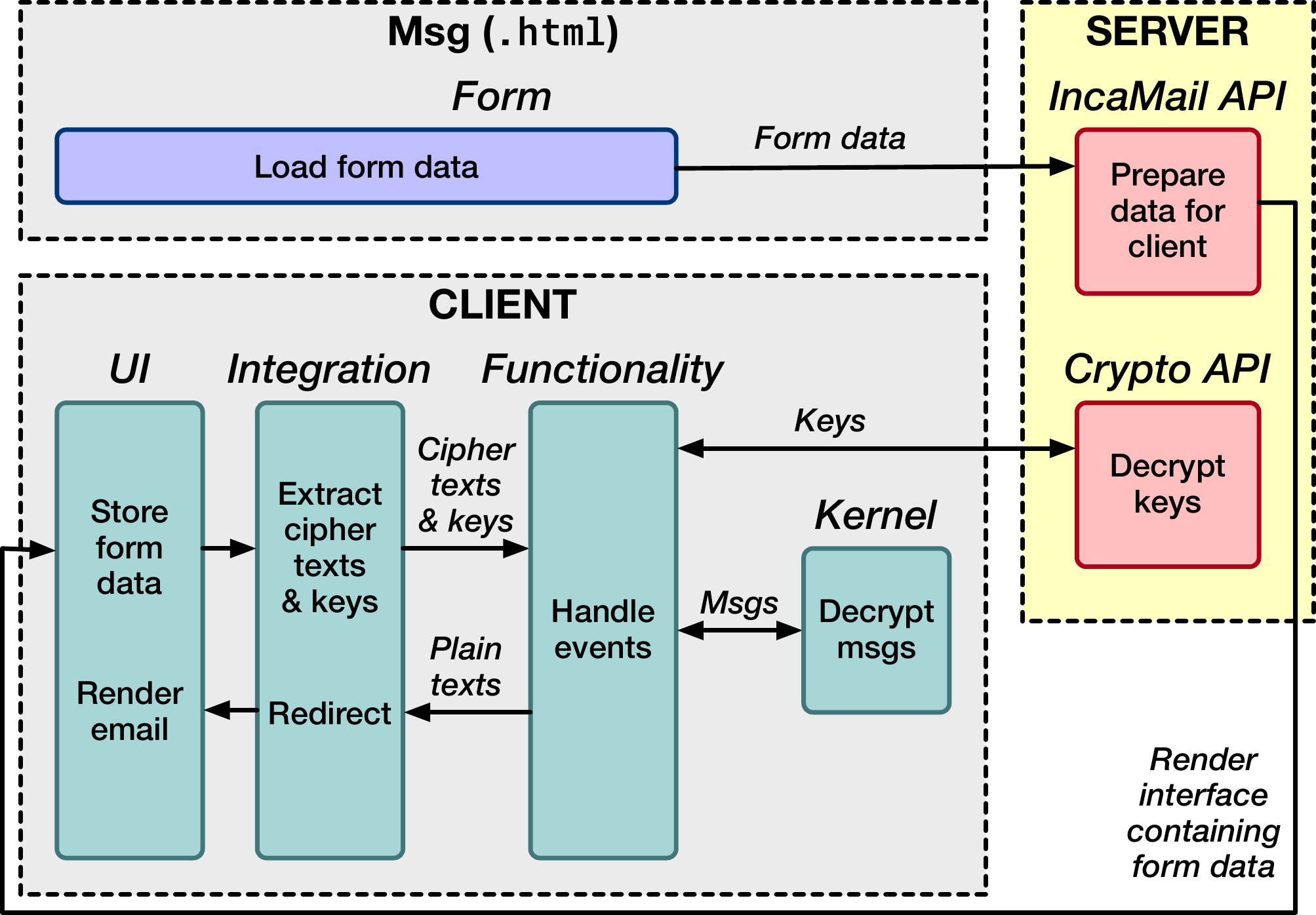}
    \caption{Workflow of reading secure messages.}
    \label{fig:reading}
\end{figure}

\begin{figure*}[t]
    \centering
    \includegraphics{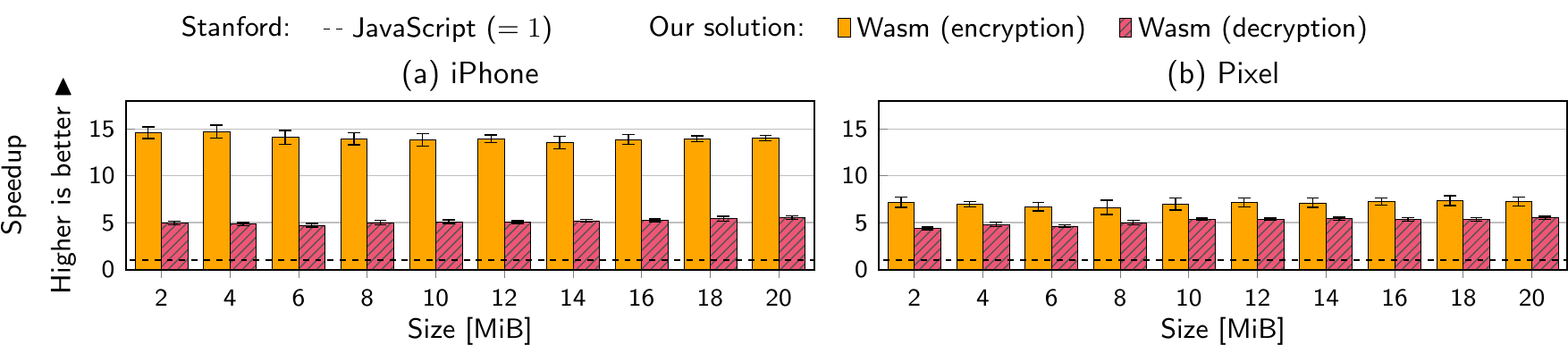}
    \vspace{-14pt}
    \caption{Speed up of the Wasm encryption and decryption, compared to a pure JavaScript implementation.}
    \label{fig:speedup}
\end{figure*}

\subsection{Proposed Workflow}
We developed a proof-of-concept prototype of the adapted architecture.
We present next the new workflows for using IncaMail with client-side encryption.

Figure \ref{fig:sending} depicts the secure message-sending workflow.
The server-side software consists of the pre-existing \emph{IncaMail API} and the newly-developed Crypto API.
The client-side software includes three layers for seamless integration with diverse clients:
\emph{(1)} the functionality layer, which contains generic event handlers invoked by the integration layer;
\emph{(2)} the kernel layer, which provides the essential cryptographic functionality for message encryption and decryption; and
\emph{(3)} the integration layer, responsible for linking the graphical user interface to the functionality layer.

The sending encryption process involves the following steps:
\emph{(1)} the user authenticates, composes a secure message, optionally appends secure files, selects recipients, and submits the form;
\emph{(2)} the kernel (\ie OpenSSL compiled in Wasm) generates an encryption key and encrypts the file attachments and the message;
\emph{(3)} the encryption key is transmitted to the Crypto API;
\emph{(4)} the ciphertexts are forwarded to the IncaMail API, which embeds them in an HTML file for assisted decryption; and
\emph{(5)} finally, the IncaMail API sends a standard email to the recipient, notifying them that the secure message is awaiting on the platform.

Figure \ref{fig:reading} illustrates the process of a recipient reading a secure message, which is described by the following steps:
\emph{(1)} the recipient receives an email containing an HTML attachment;
\emph{(2)} upon opening the HTML file, the recipient is prompted to read the secure message by clicking a button, which submits the ciphertext and associated data to the IncaMail API;
\emph{(3)} the IncaMail API then serves a page that loads the client-side software;
\emph{(4)} the client-side software requests the encryption key; and
\emph{(5)} the kernel decrypts the message and files attachments.
As the decryption is executed asynchronously on the client side, the page remains unblocked, and the file attachments are decrypted subsequent to processing the secure message. 
This enhances the responsiveness of the rendering compared to a former server-side approach. \section{Evaluation}\label{s:eval}
In this section, we present an evaluation of our proposed solution, aiming to address the following research questions:

\begin{itemize}[leftmargin=*]
    \item How does the performance of the encryption scheme implemented in Wasm compare to plain JavaScript?
    \item Does the time complexity of the encryption cipher remain linear when utilising Wasm?
\end{itemize}

To answer these questions, we employ a micro-benchmark that measures the encryption and decryption times for payloads of varying sizes (\S\ref{s:encdec}).
In \S\ref{s:complexity}, we assess whether the encryption scheme maintains linear time complexity.

\subsection{Experimental setup and methodology}
We conduct benchmarks on a range of mobile phones, including Apple iPhone Pro 12/13/14 (running iOS 16) and Google Pixel 5/Pro 6/Pro 7 (with Android 12 for the former and Android 13 for the latter two).
We leveraged the online platform LambdaTest~\cite{lambdatest} to execute the benchmarks on actual devices.
As the benchmarks solely operate on the client side, we did not consider network latency.
We note that Safari uses the WebKit~\cite{webkit} browser engine, whereas Chrome uses Blink~\cite{blink}, a fork of WebKit.
The use of different browser engines is imposed by Apple policy, forcing iOS devices to employ WebKit strictly~\cite{appleforcewebkit}.
We compiled OpenSSL v3.0.5 using Emscripten v3.1.34.

\subsection{Microbenchmark: encryption scheme}\label{s:encdec}
We devise a micro-benchmark to assess the performance of our Wasm implementation.
We compare against a pure JavaScript implementation by Stanford~\cite{5380691}, which explored various optimisation strategies for executing cryptographic operations within a JavaScript engine.
The benchmark utilised the AES cipher with a key length of \qty{128}{\bit} for encrypting and decrypting payloads generated by a pseudorandom number generator (PRNG).
Payload sizes ranged from \qty{1}{\mebi\byte} to \qty{20}{\mebi\byte}, with the maximum corresponding to IncaMail's file size limit for email attachments.
We measured the time taken by the cryptographic library to compute ciphertext and plaintext using the function \texttt{performance.now} for the JavaScript implementation, and the function \sloppy{\texttt{clock\_time\_get}} with a monotonic clock for Wasm.
The benchmark disregarded the setup and teardown time for individual runtimes.
Results were aggregated by device type, specifically \emph{iPhone} and \emph{Pixel}, due to the negligible differences observed among the models.
Consequently, we focused on the results obtained from the iPhone Pro 14 and Google Pixel Pro 7 devices for this analysis.
Each experiment was executed ten times per device, with the mean value used to determine the average outcome.

Figure \ref{fig:speedup}a and \ref{fig:speedup}b present the results obtained using iPhone and Pixel devices, respectively.
We first observe that the speedup remains consistent, irrespective of the payload size for encryption or decryption tasks.
On an iPhone, the encryption speedup using Wasm compared to JavaScript is \iphoneproFourteenEncSpeedup$\times$, while on a Pixel device, it is \pixelproSevenEncSpeedup$\times$.
Although we did not explore the precise cause of superior performance on Apple devices, it is suspected to be due to a more optimised Wasm runtime for executing OpenSSL.
Indeed, the absolute time taken by an iPhone to encrypt \qty{20}{\mebi\byte} of data using Wasm is \qty{\iphoneproFourteenEncWasmMsMaxSize}{\milli\second}, whereas on a Pixel device, it is \qty{\pixelproSevenFourteenEncWasmMsMaxSize}{\milli\second}, resulting in a ratio of \encWasmMsMaxSizeRatio\ between the two systems.
In contrast, when comparing the same operation using pure JavaScript, we observe a ratio of \encSjclMsMaxSizeRatio\ (\qty{\iphoneproFourteenEncSjclMsMaxSize}{\milli\second} on an iPhone and \qty{\pixelproSevenFourteenEncSjclMsMaxSize}{\milli\second} on a Pixel device).
The decryption speedup using Wasm is \iphoneproFourteenDecSpeedup$\times$ for both iPhone and Pixel devices.

The findings of this benchmark illustrate that Wasm significantly outperforms plain JavaScript code in cryptographic operations, thereby enabling the efficient offloading of server-side operations to Incamail's clients.

\subsection{Scalability of the encryption cipher}\label{s:complexity}

The deployment of novel technologies in constrained and restricted environments, such as in web browsers, may be subject to additional overheads and performance penalties.
We further analysed the results of our benchmark to assess the scalability of cryptographic operations, specifically examining whether the time complexity of the implementation increases with larger payload sizes.
In this experiment, we selected the execution time to encrypt and decrypt payloads ranging from \qty{1}{\mebi\byte} to \qty{20}{\mebi\byte}, subsequently dividing the obtained measurements by the time taken to encrypt \qty{1}{\mebi\byte}.

\begin{figure}
    \centering
    \includegraphics{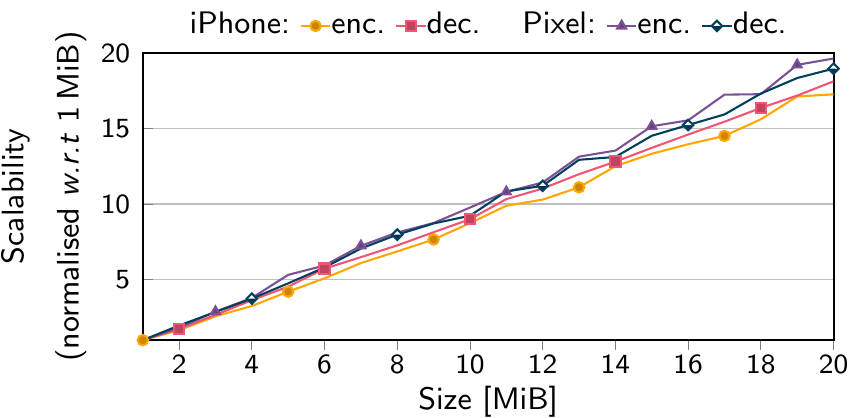}
    \vspace{-13pt}
    \caption{Scalability of the encryption cipher using Wasm, with respect to the size of the plaintext.}
    \label{fig:scalability}
\end{figure}

Figure \ref{fig:scalability} depicts the scalability of the previously-conducted benchmark for encryption and decryption operations by device type.
The results reveal that the time complexity of these cryptographic operations is linear when utilising Wasm, \ie the complexity can be expressed as $\mathcal{O}(n)$, where $n$ denotes the number of bytes to process.
Given that the execution time of the AES algorithm is linear, our strategy of offloading these cryptographic operations to the client side does not compromise the user experience, as the implementation remains efficient and performant.

 \section{Conclusion}\label{s:conclu}
This work presents a proof-of-concept that demonstrates the offloading of cryptographic operations from IncaMail's centralised architecture to clients' browsers using WebAssembly.
By implementing OpenSSL compiled in Wasm, we achieved a significant speedup in encryption and decryption tasks compared to JavaScript implementations, while maintaining a linear time complexity relative to secure message and file attachment sizes.
Furthermore, the implementation shares a common encryption key for all recipients of a given message, resulting in non-volatile memory savings.
The new architecture reduces Swiss Post's resource usage, minimises client-server traffic, and strengthens the security posture by minimising the trusted computing base.
Our findings indicate that Wasm is a viable and efficient solution for offloading cryptographic operations to the client side, improving the performance and security of secure email communication services such as IncaMail.
Future work involves integrating this proof-of-concept into a production-ready solution.
 
\textbf{Acknowledgments.}
This publication incorporates results from the VEDLIoT project, which received funding from the European Union's Horizon 2020 research and innovation programme under grant agreement number 957197.

\printbibliography
\end{document}